\documentclass[aps,prd,showpacs,floatfix,preprintnumbers,nofootinbib,amssymb,amsmath]{revtex4}
\usepackage{graphicx,bm,color}
\usepackage{subfigure}
\newcommand{\cC}{{\cal C}}
\def \etal{{\sl et al.~\/}}

\def \jhep{{\sl J.\ H.\ E.\ P.~\/}}

\def \np{{\sl Nucl.\ Phys.~\/}}

\def \pl{{\sl Phys.\ Lett.~\/}}
\def \pr{{\sl Phys.\ Rev.~\/}}

\begin{document}
\title{Astrophysics of Strange Matter}
\author{Sanjay K.\ \surname{Ghosh}}
\email{sanjay@bosemain.boseinst.ac.in}
\affiliation{Department of Physics, Bose Institute, 
          93/1, A.P.C. Road, Kolkata 700 009, India. \\
and \\
Centre for Astroparticle Physics \& Space Science, Block EN, Sector V, Salt Lake, Kolkata - 700 091, India}
\begin{abstract}
The QCD phase tansition has important consequences in the context of both the early universe as well
as compact stars. Such transitions are being studied for high temperature and small chemical potential scenario in the laboratory. There are also plans to study systems with large chemical potential and small temperatures. Here we have
reviewed the role of strange quark matter and the phase transition in all the above scenarios.
\end{abstract}
\maketitle

\section{Introduction} \label{sc.intro}

The study of strongly interacting system is an active area of research. The existence
of quark-hadron phase transition, order of transition, signature and consequences of such transitions
in different scenarios are some of the topics which are being studied both experimentally as well as 
theoretically. The fact that the QCD perturbative series shows poor convergence except for very small
coupling at very high temperatures($ \alpha_s < 0.5$ $T \sim 10^{5} T_c$), makes the calculations 
more difficult. On the other hand, the fact that nature 
has provided us with two extreme scenario in the phase diagram, makes this area of research 
more intriguing as well as interesting. 

One of the natural scenario is the large T with vanishingly small chemical potential - the scenario 
which is supposed to have existed in the Early Universe and a similar situation is supposed to arise in 
the ultrarelativistic heavy ion collisions - the central collision in the central rapidity region, 
ideal hydrodynamics and longitudinal expansion mimicking the hubble expansion.

Off course there are differences, for example, Hubble expansion time scale is
much larger than strong interaction scale whereas in ultrarelativistic collision
expansion rate is comparable to the strong interaction.

The second natural scenario is the large chemical potential and small temperature
limit which is expected to exist inside neutron stars. 
 At present we do not know whether quark matter does exists inside
the neutron stars. But with so many observational programmes going on it is really
worth to study the possibilities of exsistence of such matter in the universe and
especially inside the neutron star. This may lead us to that unique signature,
the observational detection of which will resolve our doubt.

At present there exist large number of observational data on mass-radii of neutron stars.
But, except very soft equation of states most of the other equation of states can
explain these static properties within tthe error bars \cite{schafer}. So it is necessary
to study the different dynamical features of quark matter in its various forms.

\section{Compact Stars}

The theoretical study of role of strange matter in the context of compact stars has many facets. 
For example, it is important to have a good understanding of the 
 all possible symmetry structures and the corresponding different forms of quark matter at very
high densities. It is necessary to study the consequences of the existence of
strange quark matter inside the neutron star and finally the consequences of hadron
quark phase transition itself, such as the possibility of observation of drastic change, during
phase transition, in some physical observables like breaking index. 
These consequenses will also depend on how the conversion of nuclear matter to
quark matter occurs i.e. the mechanism of phase transition itself.

\subsection{Symmetry and different quark matter phase}
At very high densities and small temperatures, because of asymptotic freedom
interaction is very weak. For higher momenta and near fermi surface quarks are almost free.
In general at zero temperature, in absense of interaction, fermi energy is same
as the chemical potential and adding or subtracting a particle would not cost anything.
So if there is an attractive potential then one can add a pair of particles with quantum number
of this attractive channel. The potential energy of this attraction will then lower
the free energy and it will become more favourable. Many such pairs will then be created near
the fermi surface. For QCD, the colour coulomb interaction is attractive between quarks having
antisymmatric colour wave function. So the pairing in colour  space is
really a natural consequence of QCD theory itself. Since pair of quarks can not be colour
singlet the cooper pair condensation in colour space will spontaneously break the color symmetry
and  gluons will acquire mass.

\begin{figure}
\includegraphics[width=5.00in]{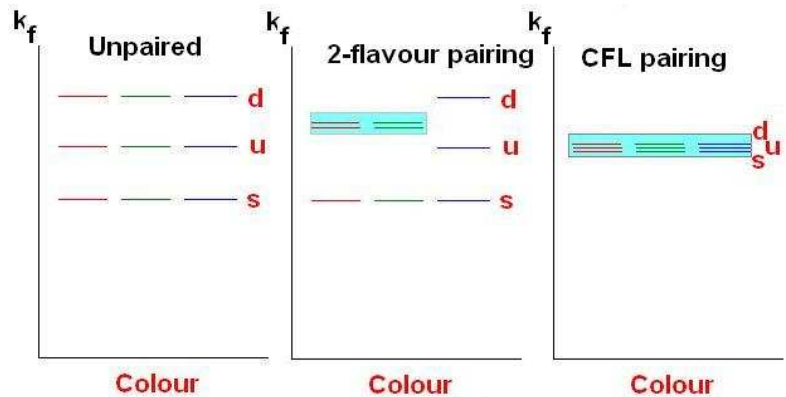}
\caption{Pairing in colour space - extreme left shows the unpaired quarks at smaller
densities or chemical potential $\mu$, with fermi momenta spacing 
$\delta{k_f} \sim \frac{{M_s}^2}{4\mu}$. For larger densities or $\mu$, 2-flavour pairing and then
3-flavour pairing or CFL phase becomes favourable.}

\label {color_fig}
\end{figure}

Again since, quarks also have flavour and spin along with colours, there
will be many variety of patterns due to quark cooper pairing. The scenario
is explained in figure \ref{color_fig}. 
In general the strange quark mass is higher
than u and d quark masses. So in matter fermi momenta of s quark is lower. From the charge
neutrlity condition number of d quarks and hence their fermi momenta is
larger. So when both neutrality and flavour equlibrium is satisfied all the
quarks tend to have different fermi momenta for the same baryon and electron
chemical potential. For small densities fermi momenta of different flavours are too far apart to have any form
of pairing and we get unpaired quarks.At higher densities, the diffference between the fermi momenta decreases. 
As a result, we initially will have pairing of u and d which have the same small mass. At even larger
densities, the fermi momenta of all the three flavours become very close and the
matter ends up in colour-flavour locked (CFL) phase, in which,  all the colour and flavour pair with each other.
This is only possible if the energy cost to make the difference in fermi momenta zero
can be compensated by the energy released due to the formation of the cooper
pairs i.e. $\Delta_{CFL} > Ms^2/ \mu$. So only for very high chemical potential
or density, such a scenario becomes possible.

As mentioned above, the stability of different forms of quark matter
with different pairing depends on the gap energy $\Delta_{CFL}$. Depending on the value of $\Delta_{CFL}$,
various form of pairing is expected to occur inside the compact star\cite{alford1}. 
There are many consequences of these
structures for the neutron star physics. For example the cooling rate is given by direct URCA
and goes
as $T^6$ for unpaired quark matter, nuclear matter at higher densities and some
other forms of matter \cite{iwamoto,ghoshneut}. On the other hand at very low densities, if proton
fraction is lower than 0.1 it is modified URCA process which works here and
cooling rate varies as $T^8$. 
In case of CFL matter, the emissivity is dominated by the Goldstone modes.
But their emissivity is suppressed by the Boltzman factor. On the other hand,
Neutrino emission from the process involving $\phi$, the massless Goldstone
boson associated with breaking of $U(1)_B$ symmetry goes as $T^{15}$ \cite{jaikumar,reddy}.
The present observational scenario seems to indicate that some neutron stars cool
much faster than others. So most probably it is the lighter nutron stars which
cool following the modified URCA process whereas the heavier nutron stars
contain some form matter which follws direct URCA process, e.g. the unpraired quark
matter or high density nuclear matter with hyperons or one of the non-CFL
superconducting phases\cite{alford1}. 

\subsection{Pulsating modes and Gravitation Wave}
The other important aspect, for which
observational efforts are being put, is the detection of gravitational waves
from the neutron stars. Gravitational waves are ripples in the space-time
curvature propagating through the space with velocity of light. Since
neutron stars are very compact objects, i.e. they have high mass within a small radius,
oscillating or more appropriately pulsating neutron stars can be very good
source of gravitational wave that could be detected by the detectors.

In general neutron stars have a large number of families of pulsating modes
with distinct characteristics. The one which is important in the present
context corresponds to bulk flows in a rotating star and is known as
Rossby mode or r-mode \cite{andersson1}. The restoring force of these is the coriolis force and
it transfers the star's angular momentum into gravitational radiation.
So for a rotating neutron star, there is a critical frequency above which
the r-mode instability sets in, angular momentum gets transfered
to gravitational wave and star spins down \cite{andersson2}.
This r-mode instability is limited by the viscous damping. For a larger viscosity
the critical spin at which r-mode becomes unstable is higher. Since the bulk viscosity of 
normal strange quark matter is larger than that of normal neutron star matter, 
an observation of newly born pulsar spinning near the Keplarian limit would provide
the evidence for a strange quark star \cite{madsen}. Similar situation is expected
for 2SC quark star as well.

In contrast, CFL matter has been shown to have very small shear damping and bulk viscosity. 
So the heating effect due to the viscous dissipation is damped in CFL stars and r-mode 
damping becomes more important for their evolution. So except for first few hundred years, 
CFL stars will cool very slowly and can exist at higher temperatures for many years \cite{zheng}.

The unstable r-modes seems to affect the strange stars differently as compared to pure (no quark matter) 
neutron stars \cite{andersson3}. Unlike neutron stars, the onset of r-mode instability, instead of leading 
to the thermo-gravitational runaway, results in the evolution of strange star to a quasiequilibrium state.
Moreover, for strange stars, r-mode instability never grow to large amplitudes.

The gravitational wave bursts induced by r-mode spin down of
hybrid stars has also been suggested. It has been proposed that continuous
emission of gravitational waves due to r-mode instability from a star can
induce a sudden variation in its structure and composition generating further
bursts of gravitational waves. This scenario is more probable for hybrid stars due to the
surface tension between the hadronic and quark matter \cite{drago1}. 

The r-mode instability and the corresponding spin down will cause an increase in
the central density of the star. This sudden increase may trigger a
phase transition inside the core of neutron star.
In strange quark matter, the strangeness fraction i.e. the ratio of
strange quark and baryon number densities will be unity if one considers
u, d and s masses to be same. Even for relativistic quark masses
($m_s > m_u \sim m_d$) the strangeness fraction at high density is not much
smaller than unity. On the contrary, the strangeness fraction in hadronic
matter is usually small. Even with hyperons the strangeness fraction is
smaller compared to quark phase. Off course with kaon condensation,
the situation may be different. But then with a kaon condensation
inside neutron stars, the transition to quark matter is found to
be pushed towards much higher densities \cite{ghosh_kaon}. So basically, the transition
from hadronic to quark matter may be associated with large strangeness production.
This can be explained in the following way. Initially hadronic matter, in terms of
quark content consist predominantly of u \& d and some s quark due to the hyperon
population. The matter is certainly out of equlibrium. The weak interaction converts
this chemically non-equilibrium matter to equilibrated matter with roughly equal
number of u, d and s quarks. This conversion is associated with the release of large
amount of energy in the form of neutrinos with average energy ~ 100 Mev \cite{ghoshnpa}.
Simple energy consideration shows that amount of energy released in agreement
with the observed energy release in gamma ray bursts.

\subsection{Effect of Phase transition}
Hadronic matter to quark matter phase transition can be studied for a rotating neutron stars. The production
of equilibrated strange quark matter proceeds through non leptonic decay 
\( u+d \leftrightarrow u+s \) and semi leptonic decays 
\( d(s) \rightarrow  u + e^{-} + \bar{\nu_e} \), \( u \rightarrow d(s)+ e^{+} + \nu_e \)
and reverse reactions. The transition from initial to final state 
will depend on the contribution of all posssible rates.
\begin{eqnarray}
\frac{dn_u(t)}{dt} = R_{d\rightarrow u}(e^-) + R_{s\rightarrow u}(e^-) - R_{u\rightarrow d}(e^-)
R_{u\rightarrow s}(e^-) + R_{d\rightarrow u}(e^+) + R_{s\rightarrow u}(e^+)
- R_{u\rightarrow d}(e^+) - R_{u\rightarrow s}(e^+)
\end{eqnarray}
where \( R_{d\rightarrow u}(e^-) \) is the reaction rate for the u quark production from d quark
via electron process. The other rates in the above equation can be defined similarly.

\begin{figure}
\includegraphics[width=5.00in]{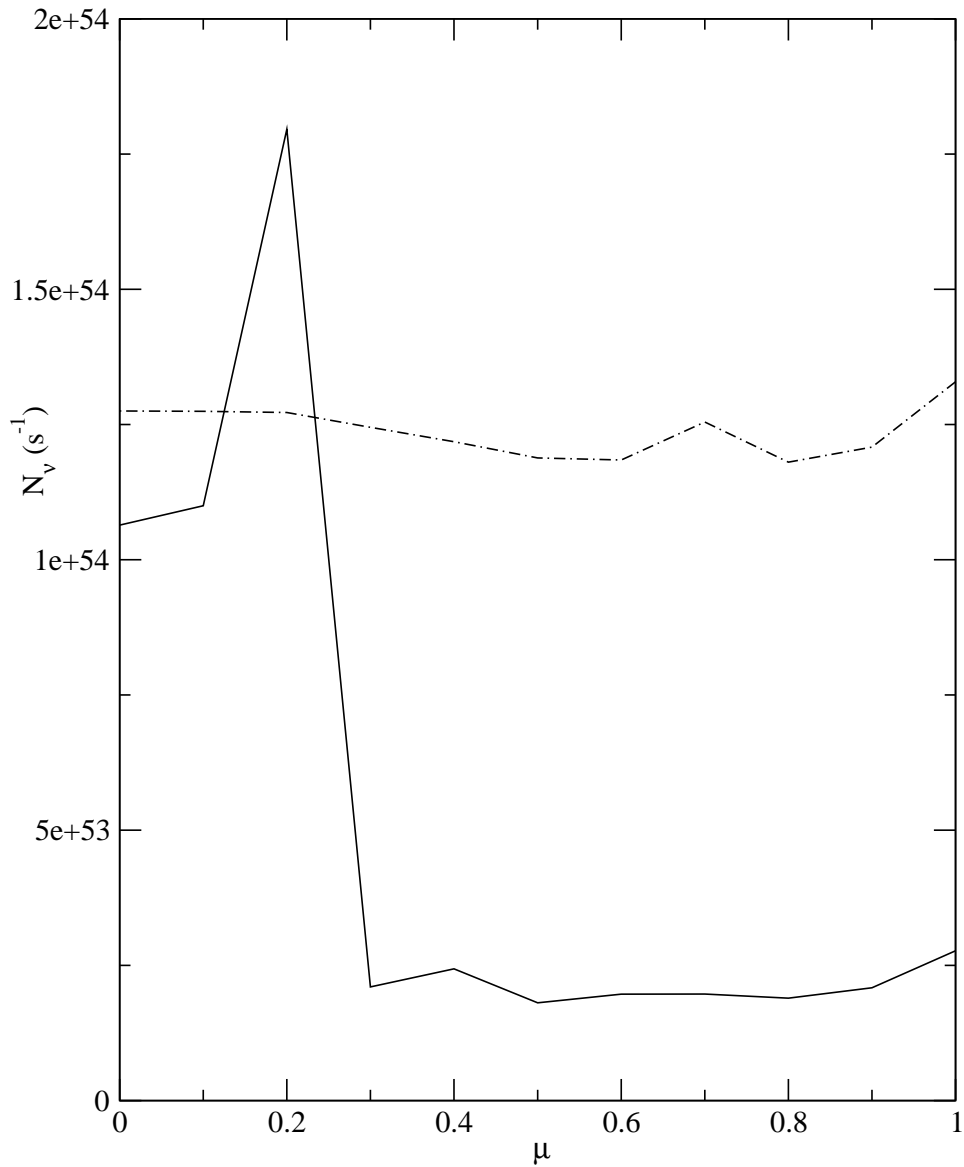}
\caption{Neutrino emission as a function of $\mu$ for two central energy densities ; 
\( 6 \times 10^{14} g/cm^3 \) (solid line) and \( 1 \times 10^{15} g/cm^3 \)
(dashed line)}
\label{neutrino1}
\end{figure}

The total number of neutrinos \( N_{\nu} \) is obtained by the volume integration over the rotating star,
\begin{eqnarray}
N_{\nu} = 2\pi \int r^2 dr n_{\nu} n_{B} \frac{e^{2\alpha + \beta}}{\sqrt{1-v^2}}
\end{eqnarray}
where \( n_{\nu} \) is the number of neutrinos emitted per unit time per baryon, \( n_{B} \)
is the baryon number density, \( \alpha \) and \( \beta \) are the gravitational potential and 
\( v \) corresponds to the rotational velocity. In figure \ref{neutrino1} we have plotted the \( N_{\nu} \)
as a function of cosine of the polar angle \( \mu ({\mathrm cos}\theta) \) for stars with two different 
central densities. For central density \( 6 \times 10^{14} gm/cm^3 \) there is a sharp peak between 
\( \mu = 0.1 \) and \( \mu = 0.24 \) with a width of \( 12^0 \). On the other hand 
for larger central density, \( 1 \times 10^{15} gm/cm^3 \),  matter concentration towards the polar region increases so that the  beaming becomes less pronounced. This indicates that for
stars with higher central density or larger quark matter region inside the compact star will show
less beaming in the neutrino spectrum \cite{ghoshplb}.

The neutrino beaming found here may be the missing link that causes the GRB. In general,
cross section for the reaction \( \nu + \bar\nu \rightarrow e^- + e^+ \) is very small. It has been shown that the general 
relativistic effects may enhance this cross section substantially and more than
 10\%  of the energy emitted in neutrinos may be deposited in \( e^+ e^- \).
This enhancement is due to path bending of the neutrinos which in turn
increases the probability of head on \( \nu \bar\nu \) collision. So it is
now necessary to calculate the energy deposition rate in the process 
\( \nu + \bar\nu \rightarrow e^- + e^+ \) for rotating compact stars. A detailed account of
the rotation along with GTR effect is given in \cite{ritam1} .
It has been shown that effect of rotation along with the general relativity enhances the energy deposition rate substantially. Hence it can provide a very efficient engine for the
gamma ray bursts.

There are some other interesting approaches as well. For example, at a temperatures of 30-60 MeV, 
neutrino emmission becomes collimated
with a beaming angle \( \theta \) around its magnetic axis, 
in the presence of surface magnetic field \( \equiv 10^{14-17} \)G. This happens in the early cooling 
evolution of strange star with colour superconducting quark matter core in the CFL
phase \cite{berdermann}.

\subsection{Mechanism of transition}
As mentioned earlier, we have really considered the phase transition
to be a two step process \cite{ritam2}. In the first step,  the nuclear matter, which is
predominantly n, p, e$^{-}$ matter gets deconfined to u , d matter i.e.
predominantly two flavour matter which then gets converted to three flavour
equilibrated matter through weak interactions. For neutron
star environment, one should really study the conversion along with 
general relativistic effect. Our study shows that the GTR effect leads to qualitative
changes in the result compared to special relativistic treatment for the
conversion of hadronic matter to 2-flavour quark matter. 
In fact the GTR effect gives rise to different conversion fonts propagating
with different velocities along
different radial directions. A detailed study of these are given in \cite{ritam3}.
The above result may have important physical consequences for the neutron star Physics.

\section{Quark Matter in Early Universe}
Let us now move over to early universe scenario. According to WMAP data
composition of the Universe is 5\% baryonic matter 25\% dark matter and 70\% dark energy.
This understanding is based on \( \Lambda \) CDM model \cite{wmap}.
The universe is expected to have undergone various changes  at different epochs
corresponding to different energy scales, starting from GUT scale of  \( E \approx 10^{19} \) GeV to
recombination at around \( E \equiv 10 \) eV. In the present paper, we will concentrate on the QCD epoch only,
at an energy scale of 100 MeV. It is believed that universe went through a phase transition from quark, gluon
system to nucleonic system. This phase transition is important as it not only sets the initial condition for 
nucleosynthesis, the relics generated at this transition might be observable today. 

Most of the relics, proposed in the literature, forms during the QCD transition only if the transition is of first order.
For example, strange quark matter as dark matter candidate, gravitational waves from colliding bubbles and
generation of magnetic field. 

Off course, a CDM candidate like QCD balls, as proposed by Zhitnitsky \cite{zhitnitsky} does not require a first order 
phase transition. Such QCD balls are formed when under certain conditions axion domain walls trap a 
large number of quarks.  This collapse is stopped by the fermi pressure of quarks which would settle 
in a colour superconducting phase.

Let us first understand the difference between the laboratory and early universe scanario. A phase transition occurs
in local thermodynamic equilibrium if the rate of equilibration which will be similar to collision rate is larger 
than the cooling rate which again is similar to expansion rate as cooling occurs through expansion. In the early universe
at the time of phase transition expansion time scale is of the order of inverse of hubble expansion 
which is much larger than the strong interaction time scale \( 10^{-23} \) seconds. This
implies that the transition, in the early universe is most likely to occcur in thermodynamic equilibrium. 
On the other hand, in ultra-relativistic heavy ion collisions, expansion rate is comparable to the 
strong interaction scale and hence the nonequilibrium effects would be important for the transition.

The recent consensus seems to be that for the physical masses of quarks (two light u and d and one heavier s quark)
there is a sharp crossover between the high temperature gas of quarks and gluon quasiparticles and 
a low temperature hadronic phase without ony thermodynamic discontinuities \cite{karsch,gupta}.

Let us now deliberate on the issue of the order of phase transition. 
In general a phase transition is said to be of order n, if the nth order derivative of the free 
energy with respect to the external
field is discontinuous or divergent. More specifically, for a second order transition, second derivative of 
free energy with respect to temperature
is discontinuous, first order derivative being continuous. Such a transition give rise to scale independent 
long wavelength fluctuations so that large regions behave coherently. On the other hand, for a first order transition, 
1st order derivative of free energy with respect to external variables is discontinuous. The value of the order parameter 
and the position of the minimum of the free energy 
jump discontinuously at the critical temperature. The difference in the free energy between the local and global minimum
yields a latent heat which is released upon the decay of the metastable state. The speed of sound 
\( {\frac{\partial P}{\partial \rho}}_T = {c_{s,T}}^2  \) jumps from zero at $T = T_c$ to a finite value
at $ T > T_c$. Here the fluctuation are scale dependent and gives rise to the formation of domains. 

\begin{figure}[!tbh]
   \includegraphics{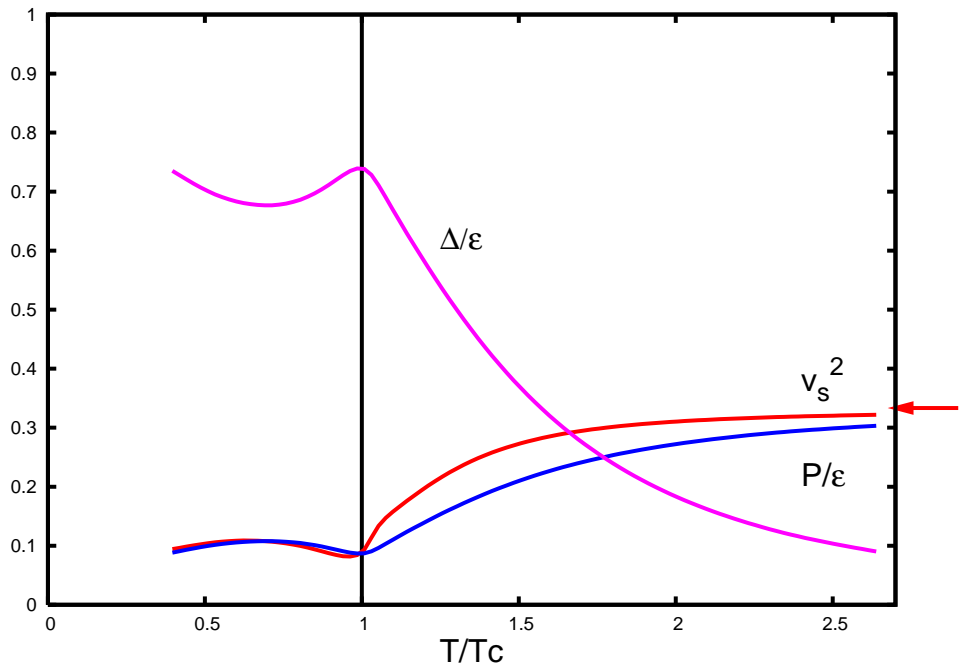}
   \caption{Squared velocity of sound $v_s^2$ and conformal measure
           $\cC=\Delta/\epsilon$ as function of $T/T_c$. The arrow on
           the right shows the ideal gas value for $v_s^2$. For 
           comparison with $v_s^2$ we also plot the ratio $P/\epsilon$.
           }
\label{fg.ccs}\end{figure}

Recent lattice calculations  of QCD transition \cite{aoki} have shown a sharp drop in the
speed of sound as temperature decreases towards the critical temperature. This implies that the 
system going through a cross over transition may behave somewhat similar to 
first order transition. In figure \ref{fg.ccs} we have plotted the speed of sound along with the conformal measure
\cite{tamalprd} calculated form PNJL model \cite{pnjl}. 
The $v_s^2$ is close to its ideal gas value at the temperature of about
$2.5 T_c$. This is close to the results for pure glue theory on the
Lattice as reported in Ref.\cite{swa2} and also that with 2 flavour 
Wilson Fermions in Ref.\cite{khan}, with 2+1 flavours of staggered 
quarks reported in Ref.\cite{szabo}, and with improved 2 flavour 
staggered fermions ($P/\epsilon$ was measured in this case).  However, 
near $T_c$ the $v_s^2$ in Ref.\cite{swa2} goes to a minimum value above
0.15, whereas we find the minima going close to 0.08, consistent with 
simulations with dynamical quark in Refs.\cite{khan,szabo}, and 
remarkably close to the softest point $P/\epsilon = 0.075$ in 
Ref.\cite{isentropp}.

In case of a 1st order transition, one would expect the fromation of trapped false vacuum domains (TFVD) during the transition
\cite{witten} some of which may become stable depending on their baryon number content \cite{alam,abhijit_alam}.
These quark nuggets may further evolve giving rise to massive compact obejects \cite{banerjee_mnras}. 
The above process gives rise to an interesting phenomena when one considers the role of colour charges 
explicitly \cite{banerjeeplb}.

Let us assume the colour wave function of the universe to be colour singlet prior to the QCD phase transition,
or in other words, wave function of all the coloured objects are completely entangled \cite{wootters}.
In such a situation the universe is characterized by the vacuum energy of perturbative QCD. During the
transition, colour neutral configurations (Hadrons) arise which results in the gradual decoherence of the 
entangled colour wave function of the entire universe. This amounts to aproportionate reduction in 
the perturbative vacuum energy density which goes into providing the latent heat of the transition. 
The TFVDs formed during the transition would try to release the residual colours to become colour neutral. 
So at the end of the transition one would be left with coloured charges separated by space like distances.
From the quantum mechanical point of view, the colour wave function of the largely separated quarks
still remains entangled  and the corresponding amount of perturbative energy would persist \cite{banerjeeplb}
which would then play the role of cosmological constant.
Now, for a cross over transition, TFVDs may or may not evole to  stable quark nuggets. But the existence of
scale dependent fluctuations is enough for the the formation of TFVDs and the quarks separated by space like distance. 
In other words, even for a cros over  transition, the QCD transition may provide a solution to the
dark energy problem.

\section{summary}
QCD transition is one of the mose interesting areas of research. We have discussed the impact of the
QCD transition in natural scenario of neutron stars and early universe. Though, an observational signature
of phase transition and strange quark matter in nature is still elusive, the study of these phenomena
is extremely important for a better understanding of the physics of strongly interacting matter. Moreover, there are 
many questions remain to be answered before we can make any conclusions.


\end{document}